\documentclass[a4paper,12pt]{article}
\usepackage{amsmath,amssymb,amsfonts}
\usepackage[utf8]{inputenc}
\usepackage[english,russian]{babel}
\usepackage[normalem]{ulem}  
\usepackage[]{fontenc}
\usepackage{pst-circ}
\usepackage{mathrsfs}
\usepackage{tasks}
\usepackage{amsmath,amssymb,amsfonts}
\usepackage{breqn}
\usepackage{pb-diagram}
\def\around{\circlearrowleft}

\def\C{\mathbb{C}}
\def\WW{U}
\def\d{\partial}
\def\D{\mathcal D}
\def\F{\hat{\Lambda }}

  \def\As{\Lambda^{(-)}}
    
    \def\Sym{\Lambda^{(+)}}
 \def\po{N}
 \def\Z{\mathbb{Z}}
 \def\E{\mathcal{E}}
 \def\Fo{\mathcal{F}}
 
 \def\H{\mathscr{H}}

\def\-{\,\text{--}\,}

\newtheorem{Lemma}{Lemma}

\newtheorem{Prop}{Proposition}
\newtheorem{Remark}{Remark}

\newenvironment{Proof}{\par\noindent{\bf Proof.}}{\hfill$\scriptstyle\blacksquare$}

\begin{document}
\bigskip
\hfill{ITEP-TH-29/18}
\begin{center}
{\Large\bf Fermionic limit of the Calogero-Sutherland system}

\bigskip
{\bf S.M. Khoroshkin$^{\star\circ}$,  \, M.G. Matushko$^{\circ\ast}$
}\medskip\\
$^\star${\it Institute for Theoretical and Experimental Physics, Moscow, Russia;}\smallskip\\ 
$^\circ${\it National Research University Higher School of Economics, Moscow, Russia;}\smallskip\\
$^\ast${\it  Center for Advanced Studies, Skoltech, Moscow, Russia}
\end{center}
\selectlanguage{english}
\begin{abstract}\noindent
We present a construction of an integrable model as a projective type limit of 
Calogero-Sutherland models of N fermionic particles, when $N$ tends to infinity. Explicit formulas for limits of Dunkl operators and of commuting Hamiltonians by means of vertex operators are given. 
\end{abstract}
\section{Introduction}
{ Effective} and rigorous constructions of limits of quantum Calogero-Sutherland (CS) systems have attracted the attention of mathematicians for many years \cite{AJ,AMOS,AMY,Pol}.
Note here the fundamental research of D.Uglov \cite{Uglov}, 
where he defined and studied an  inductive
limit of  fermionic CS system. His construction waited for more than 15 years for a further development until M. Nazarov and E.Sklyanin
suggested a precise construction of higher Hamiltonians  for  the scalar CS system using the Sekiguchi determinant 
and  the machinery of symmetric functions \cite{NazSk1}. In \cite{SerVes2} A.Veselov and A.Sergeev  suggested to define the bosonic
limit of the CS system as a projective limit of finite  models. Precise bosonic constructions of higher Hamiltonians in a Fock space 
were then presented by M.Nazarov and E.Sklyanin in \cite{NazSk} and by  A.Veselov and A.Sergeev in \cite{SerVes}. The crucial point of their constructions is the use of  equivariant family of Heckman--Dunkl operators  as a quantum $L$-operator for the CS system.

This paper can be regarded as a further development of the latter ideas to the CS systems restricted to the antisymmetric wave functions. 
The both approaches \cite{NazSk}, \cite{SerVes} in the bosonic case regard the space  $\C [x_1] \otimes \Sym[x_2, \dots , x_{N}]$ of functions as a domain of action of quantum $L$-operator, 
which is effectively coincides with Dunkl operators. The space  $\C [x_1] \otimes \Sym[x_2, \dots , x_{N}]$ consists of polynomials, symmetric in all variables except one and is invariant 
under the action of the Dunkl operator $D_1$. In the limit the action of the Dunkl operator 
is defined in the space $\C[z]\otimes \F$ with a help of operators 
$$
	V_+(z)=\exp\sum_{n\ge0}z^n\frac{\partial}{\partial p_n} \qquad\text{and}\qquad \varphi_-(z)=
	\sum_{n>0}\frac{p_n}{z^n}.$$
Here $\F$ is  as an irreducible representation of the Heisenberg algebra generated by $p_n$
and $\frac{\partial}{\partial p_n}$, $n=0,1,...$ of bosonic creation and  annihilation operators respectively. 

In this paper we realize the fermionic limit for the CS system.
 
As well as in bosonic case we begin with the description of the CS system restricted to the space of antisymmetric polynomials $\As[x_1,x_2, \dots , x_N]$ in terms of Heckman--Dunkl operators. We then express Heckman--Dunkl operators via finite analogs $V_-(z)V_+(z)$ and $V'_-(z)V'_+(z)$ of vertex operators 
 $\Psi(z)$ and $\Psi^*(z)$, where
$$
\Psi(z)=z^{p_0}\exp\left(-\sum_{n>0}\frac{p_n}{n z^n}\right)\exp\left(\sum_{n\ge0}z^n\frac{\partial}{\partial p_n}\right).
$$  
 To do this we present any antisymmetric polynomial in $N$ variables as
$$
\prod_{i>j}(x_i-x_j)f(p^{(N)}_1,p^{(N)}_2,p^{(N)}_3,\dots)
$$
where $p^{(N)}_k=x_1^k+\ldots+x_N^k$. The operator $V_+(x_1)$ changes each occurrence
 of $p^{(N)}_k$ by $p^{(N-1)}_k+x_1^k$, while the operator $$V_-(x_1)=x_1^{\po}\exp\left(-\sum_{n>0}\frac{p^{(N-1)}_n}{n x_1^n}\right)$$ is the multiplication by  $ \prod_{i=2}^{N}(x_1-x_i),$
 so that the application of  $V_-(x_1)V_+(x_1)$ to an antisymmetric polynomial $g(x_1,...,x_N)$ is just its Taylor decomposition with respect to $x_1$. On the other hand, the operators  $V'_-(z)V'_+(z)$ are used for the total antisymmetrization of the functions, antisymmetric with respect to all variables except one. This is done in Section \ref{polphase}.
 
 Let  $\F=\Lambda[p_0]$ be a ring symmetric functions \cite{Mac} extended by a free variable $p_0$. 
 The space $\F$ is an irreducible representation of the Heisenberg algebra, generated by the elements $p_n$ and $\dfrac{\partial}{\partial p_n}$ and can be regarded as a polynomial version of the Fock space. It contains the vacum vector
 $|0\rangle$, such that $$\dfrac{\partial}{\partial p_n}|0\rangle=0,\qquad n=0,1, \ldots .$$
 The dual vacuum vector $\langle 0|$ satisfies the condition
 $$\langle 0| p_n=0, \qquad n=0,1,\ldots .$$
 To each vector $|v\rangle$ of $\F$ we attach a family $\{\pi_N(v)\}$ of antisymmentric functions of $N$ variables, given by matrix elements
  \begin{equation}\label {i0}\pi_N(v)=
  \langle 0|\Psi(x_N)\cdots \Psi(x_1)|v\rangle.
  \end{equation}
 The goal is to construct operators in the space $\F$ which are compatible with finite CS Hamiltonians with respect to evaluation maps (\ref{i0}). 
 This is done following E.Sklyanin ideology \cite{NazSk, KMS}: we introduce an auxillary space $U\subset \C[z,z^{-1}]]\otimes \Fo$ and its evaluations to the spaces of  polynomials antisymmetric with respect 
 to all variables except one. We present operators, acting  in $U$ which are compatible with the above evaluation maps. They are limits of Heckman--Dunkl operators, and the limiting Hamiltonians are then
 constructed by means of certain integral average of them.
  The constructed operators form a commutative family of operators in the space $\F$.
 
 Contrary to the ring of symmetric functions, the space $\F$ is not the projective limit of the spaces of (anti)symmetric functions due to the presence of zero mode $p_0$. 
   On the other hand, CS Hamiltonians $\bar{H}_k$ theirselves do not form a projective family since they do not respect natural projections
 $\lambda_N:\As[x_1,x_2, \dots , x_{N+1}]\rightarrow\As[x_1,x_2, \dots , x_N]$, 
 that is $
 \lambda_N \bar{H}_k^{(N+1)}\neq \bar{H}_k^{(N)}\lambda_N.
 $
However, the Hamiltonians $\bar{H}_k^{(N)}$ written in the form (\ref{Hpol}) are compatible with the maps $\lambda_N$, once we replace each occurrence of $N$ in $\bar{H}_k^{(N)}$
 to $N+1$ in $\bar{H}_k^{(N+1)}$. Moreover, each finite Hamiltonian can be  restored from its limit by formal replacement of each occurrence of $p_0$ by operator of multiplication on the number $N$ of particles.
 
 The constructed Hamiltonians form a commutative family of operators in the space $\F$. Moreover, they commute inside the Heisenberg algebra and thus can be used as well 
 in its other representations, for instance, in the bosonic Fock space. We can define the projection $\tilde{\pi}_N:\Fo\rightarrow \As[x_1,x_2, \dots , x_N]$ similar to (\ref{i0}) 
  \begin{equation}\notag\tilde{\pi}_N(v)=
  \langle 0|\Psi(x_N)\cdots \Psi(x_1)|v\rangle.
  \end{equation}
  In fact it is nonzero only on the $N$-th sector $\Fo_N$ of the Fock space. Now the constructed Hamiltonians $\H_k$ are compatible with respect to the maps $\tilde{\pi}_N$, the commutativity $\tilde{\pi}_N\H_k=\bar{H}_k^{(N)}\tilde{\pi}_N$
  is nontrivial on the $N$-th sector $\Fo_N$. We reformulate the same construction in the fermionic Fock space represented as space of semi-infinite wedges, we define the projection analogous to $\tilde{\pi}_N$ which acts as a ``cutting'' of the wedge.
 We discuss this in Section 5.
 
 A different approach to the construction of the limiting systen is presented in \cite{KM} in a more general context of the spin CS model.

\section{CS model}
Consider the quantum Calogero-Sutherland model of $N$ particles on the circle \cite{B,KK}. Its Hamiltonian is
\begin{equation*}
H^{CS}= -\sum_{i=1}^{N}\left(\frac{\partial}{\partial q_i}\right)^2+2\left(\frac{\pi}{L}\right)^2\sum_{i<j}^N\frac{\beta(\beta-{K_{ij}})}{\sin^2\left(\frac{\pi}{L}(q_i-q_j)\right)},
\end{equation*}
where $K_{ij}$ is the coordinate exchange operator of particles $i$ and $j$.
After conjugating by the function $\prod_{i<j}|\sin(\frac{\pi}{L}(q_i-q_j))|^\beta$ which represents the vacuum state with eigenenergy $E_0=\left(\pi\beta/L\right)^2 N(N^2-1)/3$, and passing to the exponential variables $x_i=e^{\frac{2\pi i q_i}{L}}$ we come to the Hamiltonian  
\begin{equation}\label{H}
H=\sum_{i=1}^{N}\left(x_i \frac{\partial}{\partial x_i}\right)^2+\beta\sum_{i<j}\frac{x_i+x_j}{x_i-x_j}\left(x_i\frac{\partial}{\partial x_i}-x_j\frac{\partial}{\partial x_j}\right)-2\beta\sum_{i<j}\frac{x_i x_j}{\left(x_i-x_j\right)^2}\left(1-K_{ij}\right).
\end{equation}
We consider the antisymmetric wave functions of the Hamiltonian (\ref{H}):
$$
\phi(x_1,\dots ,x_i,\dots,x_j,\dots,x_N)=-\phi(x_1,\dots ,x_j,\dots,x_i,\dots,x_N), 
$$
then the eigenfunctions of the Hamiltonian $H^{CS}$ 
$$
\prod_{i<j}|\sin(q_i-q_j)|^\beta\phi(e^{2\pi i q_1},\dots,e^{2\pi i q_N})
$$
are also antisymmetric by the variables $\{q_i\}$ except for the vacuum state. We can write the restriction of the Hamiltonian (\ref{H})  on the space of antisymmetric functions by the following formula 
\begin{equation}\label{H2}
\bar{H}=\sum_{i=1}^{N}\left(x_i \frac{\partial}{\partial x_i}\right)^2+\beta\sum_{i<j}\frac{x_i+x_j}{x_i-x_j}\left(x_i\frac{\partial}{\partial x_i}-x_j\frac{\partial}{\partial x_j}\right)-4\beta\sum_{i<j}\frac{x_i x_j}{\left(x_i-x_j\right)^2}.
\end{equation}
Further we use the Heckman--Dunkl operators $D_i^{(N)}$ in the form suggested by Polychronakos \cite{Pol,Dunkl}:
\begin{equation} \label{Dunkl}
\D_{i}^{(N)}=x_i\frac{\partial}{\partial x_i}+\beta\sum_{j\neq i}\frac{x_i}{x_i-x_j}\left( 1 - K_{ij}\right).
\end{equation}
These operators  satisfy the relations
\begin{align}\nonumber
K_{ij}\D_i^{(N)}&=\D_j^{(N)} K_{ij}, \\ \nonumber
[\D_i^{(N)},\D_j^{(N)}]&=\beta (\D_j^{(N)}-\D_i^{(N)})K_{ij},
\end{align}
which coincide with the relations of the degenerate affine Hecke algebra after the renormalization 
${\D_i^{(N)}}=\beta\tilde{\D}_i$. We introduce the operators
$$
\bar{H}_k=\mathrm{Res}_-\left(\sum_i \left(\D_{i}^{(N)}\right)^k\right),
$$
where $\mathrm{Res}_-$ means the restriction on the space of antisymmetric functions. As an exapmle
$$
\bar{H}_1=\mathrm{Res}_-\left(\sum_i \D_{i}^{(N)}\right) =\sum_{i=1}^{N}\left( x_i\frac{\d}{\d x_i}\right) +\beta N(N-1).
$$
These operators commute \cite{Heck}. Moreover,  operators $\bar{H}_k$ represent integrals of motion of the quantum Calogero Sutherland model. 
Here is the expression of the Hamiltonian (\ref{H2}) in terms of $\bar{H}_k$:
\begin{equation}\label{prH}
\bar{H}=\bar{H}_2-2\beta (N-1)\bar{H}_1 +\beta^2N(N-1)^2.
\end{equation}
\section{Polynomial phase space} \label{polphase}
\textbf{1. }We regard the CS system of $N$ fermionic particles with polynomial wave functions using the Heckman-Dunkl operators.
The corresponding Heckman--Dunkl operators $\D_i^{(N)}:\C[x_1,\dots,x_N]\rightarrow \C[x_1,\dots,x_N] $ are defined by the relation (\ref{Dunkl}).
Symmetric polynomials in $\D_i^{(N)}$ preserve the space of symmetric  $\Sym[x_1,\dots,x_N]$ and  antisymmetric polynomials $\As[x_1,\dots,x_N].$ Denote by $\alpha_N: \Sym[x_1,\dots,x_N]\rightarrow \As[x_1,\dots,x_N]$ the canonical isomorphism
 \begin{equation}\label{al}
  \alpha_N: f(x_1,\dots x_N)\rightarrow \bar{f}(x_1,\dots x_N)=f(x_1,\dots x_N)\Delta(x_1,\dots ,x_N),
 \end{equation}
where $$\Delta(x_1,\dots ,x_N)=\det_{i,j=1\dots N}(x_i^{N-j})=\prod_{i<j}(x_i-x_j)$$ is the Vandermonde determinant.

The space $\Sym[x_1,\dots,x_N]$ is generated by the Newton polynomials $p_k^{(N)}=x_1^k+\dots +x_N^k$,  $k = 1,\dots N$. (sometimes for brevity we omit the upper index $N$ and simply write $p_k$). Due to (\ref{al}) any antisymmetric polynomial can be written by the following formula
$$
\bar{f}(x_1\dots x_N)=\Delta(x_1,\dots ,x_N) f(\{p_k^{(N)}\}), \qquad k=1,2,\ldots
$$
where $f$ is a polynomial in $p_k$.  Here and further we denote by $f(x_1,\dots x_N)$ or $ f(\{p_k^{(N)}\}$ a symmetric function and by $\bar{f}(x_1,\dots x_N)$ the corresponding antisymmetric function following (\ref{al}). For 
an operator $A$ acting on the symmetric functions we denote by $\bar{A}$ the corresponding operator acting on the  antisymmetric functions so that the relation $
\bar{A}\bar{f}(x_1,\dots x_N)=\overline{Af(x_1,\dots x_N)}
$ holds.

The Dunkl operator $\D_i^{(N)}$ preserves the antisymmetry involving all variables other than $x_i$.  Denote by $\bar{D}_i^{(N)}$ the restriction of  $\D_i^{(N)}$ to the space of functions 
\begin{equation}\label{111}
\bar{f}(x_i;x_1,\dots, x_N)\in \C[x_i]\otimes \As[x_1,\dots x_{i-1},x_{i+1},\dots x_N]
\end{equation}
antisymmetric in all variables other than $x_i$.
Due to  (\ref{al}) the LHS of (\ref{111}) can be presented as
\begin{equation}\notag
\bar{f}(x_i;x_1,\dots, x_N)=\Delta(x_1,\dots x_{i-1},x_{i+1},\dots ,x_N) f(x_i;\{p_k\} ),
\end{equation}
where $f$ is a polynomial in $x_i$ and in $p_k$, which depend on $N-1$ variables.
\medskip

\textbf{2. }In the following we use the notations
\begin{equation}\label{V+}
	V_+(z)=\exp\left(\sum_{n> 0}z^n\frac{\partial}{\partial p_n}\right),\qquad
	V_-(z)=z^{\po}\exp\left(-\sum_{n>0}\frac{p_n}{n z^n}\right),
\end{equation}  
where $N$ is the number of variables in $p_k$.
More precisely, the operator $V_+(z)$ maps a polynomial expression  in $\{p_k\}$ and in $z$ into the same expression changing each occurrence of a Newton sum $p_k^{(N)}$ by $p_k^{(N-1)}+z^k$ due to the Taylor formula.
The operator $V_-(z)$ does not change the number of variables in $p_k=p_k^{(N)}$ and can be equivalently written as an operator of multiplication by $\displaystyle\prod_i (z-x_i)\in \C[z]\otimes \Sym[x_1,\dots,x_N]$:
\begin{equation}\label{V-e}
V_-(z)=z^{\po}\prod_{i=1}^{\po}\exp\left(-\sum_{n>0}\frac{x_i^n}{n z^n}\right)=z^{N}\prod_{i=1}^N\exp\left(\ln\left(1-\frac{x_i}{ z}\right)\right)=
\prod_{i=1}^N (z-x_i).
\end{equation}
 Note that further we mostly use  the  composition of operators  $V_-(z)V_+(z)$, which  maps the space $\Sym[z,x_2,\dots,x_N]$ to $\C[z]\otimes \Sym[x_2,\dots,x_{N}]$. 
In this composition the operator $V_-(z)$ has the form $V_-(z)=z^{N-1}\exp\left(-\sum_{n>0}\frac{p_n}{n z^n}\right)$,
where $p_k$ depend on $N-1$ variables.
\medskip

\textbf{3. } Let $f(\{p_k\})$ be a symmetric polynomial in $N$ variables and
$$\bar{f}(x_1\dots x_N)=\Delta(x_1,\dots ,x_N) f(\{p_k\})$$
the corresponding antisymmetric polynomial. Denote by 
$$\bar{\iota}_{N,i}:\As[x_1,\dots,x_N]:\rightarrow\C[x_i]\otimes \As[x_1,\dots x_{i-1},x_{i+1},\dots x_N]$$ the natural embedding representing any antisymmetric polynomial as
 a polynomial in $x_i$ with coefficients in $\As[x_1,\dots x_{i-1},x_{i+1},\dots x_N]$.
\begin{Prop} \label{lemma1}
	The embedding $\bar{\iota}_{N,i}$ is given by the following relation:
	\begin{equation}\label{bariot}
	\begin{split}
		\bar{\iota}_{N,i}(\bar{f}(x_1\dots x_N))& =(-1)^{i+1}\overline{\iota_{N,i}f(\{p_k\})}=(-1)^{i+1}\overline{V_-(x_i)V_+(x_i)f(\{p_k\})}=\\&=(-1)^{i+1}\Delta(x_1,\dots x_{i-1},x_{i+1},\dots ,x_N)V_-(x_i)V_+(x_i)f(\{p_k\}).
		\end{split}
	\end{equation}
	\end{Prop}
	Here $V_-(x_i)V_+(x_i)f(\{p_k\})$ is a polynomial in $x_i$ and in Newton polynomials  $\{p_k\}$ depending on $(N-1)$ variables.
\begin{Proof}
 Using the definition of $\bar{\iota}_{N,i}$ we present the antisymmetric function $\bar{f}(x_1\dots x_N)$ in the following form 
 \begin{equation}\label{gd}\begin{split}
\bar{\iota}_{N,i}(\bar{f}(x_1\dots x_N))=&\bar{f}_0(x_1,\dots x_{i-1},x_{i+1},\dots x_{N})+\bar{f}_1(x_1,\dots x_{i-1},x_{i+1},\dots x_{N})x_i+ \\
& +\bar{f}_2(x_1,\dots x_{i-1},x_{i+1},\dots x_{N})x_i^2+\dots,
\end{split}\end{equation}
where each $\bar{f}_l(x_1,\dots x_{i-1},x_{i+1},\dots x_{N})$ is an antisymmetric polynomial. The decomposition (\ref{gd}) consists of two steps. The first one is a substitution
\begin{equation*}
p_n^{(N)} \rightarrow p_n^{(N-1)}+x_i^n
\end{equation*}
in all the functions $f(\{p_k\})$, which is performed by the Taylor expansion
$$
f(z+t)=\exp\left(t\frac{\partial}{\partial z}\right){ f(z)}=f(z)+f'(z)t+\frac{1}{2}f''(z)t^2+\dots
$$
giving a finite sum for polynomials.
The second step is a factorization of the Vandermonde determinant:
\begin{equation*}
\Delta(x_1,\dots ,x_N)=\Delta(x_1,\dots x_{i-1},x_{i+1},\dots ,x_N)(-1)^{i+1}\prod_{j\neq i}(x_i-x_j).
\end{equation*}
Due to (\ref{V-e}) the factor $\prod_{j\neq i}(x_i-x_j)$ can be implemented in terms of $p_k$  by applying the operator $V_-(x_i)$. Thus we obtain (\ref{bariot}).
\end{Proof}

Observe that the formula (\ref{bariot}) is correct for any expression of the symmetric function in terms of Newton polynomials $p_k$ irrespective of their dependencies. Indeed, 
$$
V_-(z)=\prod_{i=1}^N (z-x_i)=\sum_{k \ge 0}^N e_k(x_1,\dots,x_N) z^{k},
$$
where $\displaystyle e_k(x_1,\dots,x_N)=\!\!\!\sum_{1\leq i_1< \dots< i_k\leq N}x_{i_1}x_{i_2}\dots x_{i_k}$ are the elementary symmetric polynomials. They can be expressed by Newton sums $p_k(x_1,\dots,x_N)$ using Newton identities, 
and these expressions do not depend on the number of variables $N$.
\medskip

\textbf{4. }We also use the notations
\begin{equation}\label{V+'}
	V'_+(z)=\exp\left(-\sum_{n>0}z^n\frac{\partial}{\partial p_n}\right),\qquad 
	V'_-(z)=z^{-\po}\exp\left(\sum_{n>0}\frac{p_n}{n z^n}\right).
\end{equation}  
By definition the operator $V'_+(z)$ changes each occurrence of the formal variable \\
$p_k^{(N-1)}(x_1,\dots ,x_{N-1})$
by the difference $p_k^{(N)}(x_1,\dots,x_{N-1},z)-z^k$. Thus the operator $V'_{{ +}}(z)$ maps the space  $\C[z]\otimes \Sym[x_1,\dots,x_{N-1}]$ into itself, changing the meaning of the variables $p_k$.  The operator $V'_-(z)$ can be equivalently written 
\begin{equation*}
V'_-(z)=z^{-\po}\prod_{i=1}^N\exp\left(\sum_{n>0}\frac{x_i^n}{n z^n}\right)=z^{-\po}\prod_{i=1}^N\exp\left(-\ln\left(1-\frac{x_i}{ z}\right)\right)=z^{-N}\prod_i \frac{1}{(1-\frac{x_i}{z})}=
\end{equation*}
$$
=z^{-N}\sum_{k\ge 0}\left(\sum_{1\le i_1\le i_2\le\dots\le i_k\le N}x_{i_1}x_{i_2}\dots x_{i_N}\right) z^{-k}=\sum_{k \ge 0} h_k(x_1,\dots,x_N) z^{-k-N},
$$
where $\displaystyle h_k(x_1,\dots,x_N)=\!\!\sum_{1\le i_1\le i_2\le\dots\le i_k\le N}x_{i_1}x_{i_2}\dots x_{i_N}$ are complete homogeneous symmetric polynomials.
We  then can rewrite
\begin{equation}\label{V-f}
V'_-(z)=\sum_{k \ge 0} h_k(\{p_n\}) z^{-k-N},
\end{equation}
where $ h_k(\{p_n\})$ means that complete homogeneous symmetric polynomials are expressed from the basis of the Newton polynomials. These expressions do not depend on the number of variables $N$.
So the operator $V'_-(z)$ transforms the space of polynomials in $p_k^{(N)}$ and in $z$ into Laurent series in $z$ with coefficients being polynomials in $p_k^{(N)}$.
\medskip

\textbf{5. }Acting on antisymmetric function  in $N$ variables the Dunkl operators produce an equivariant family of $N$ functions
  $$\bar{f}_1(x_1;x_2,...,x_N),\qquad \bar{f}_2(x_2;x_1,x_3,...,x_N),\qquad \bar{f}_N(x_N;x_1,...,x_{N-1}), 
  $$
 where $\bar{f}_i(x_i;x_1,\dots, x_N)\in \C[x_i]\otimes \As[x_1,\dots x_{i-1},x_{i+1},\dots x_N] $
 and  $K_{ij}\bar{f}_j(x_j;x_1,\dots, x_N)=-\bar{f}_i(x_i;x_1,\dots, x_N)$.
   
	For any polynomial $\bar{f}(x_i;x_1,\dots, x_N)\in\C[x_i]\otimes \As[x_1,\dots x_{i-1},x_{i+1},\dots x_N]$ denote by  $\bar{\E}_N \bar{f}\in\As[x_1,\dots ,x_N]$ the sum
 \begin{equation*}
 \begin{split}
 (\bar{\E}_N \bar{f})(x_1,...,x_N)=&\bar{f}(x_1;x_2,...,x_N)-\bar{f}(x_2;x_1,x_3,...,x_N)-\dots -\bar{f}(x_N;x_1,...,x_{N-1}),
 \end{split}
 \end{equation*}
 which we call the total antisymmetrization of the function $\bar{f}_i(x_i;x_1,\dots, x_N)$.
 
 Let $f(x_i;\{p_k\})\in \C[x_i]\otimes \Sym[x_1,\dots x_{i-1},x_{i+1},\dots x_N$] and $\bar{f}(x_i;x_1,\dots, x_N)$ 
 be the corresponding element of the space $ \C[x_i]\otimes \As[x_1,\dots x_{i-1},x_{i+1},\dots x_N]$:
  \begin{equation}\label{SK1}\bar{f}(x_i;x_1,\dots, x_N)=(-1)^{i+1}\Delta(x_1,\dots x_{i-1},x_{i+1},\dots ,x_N) f(x_i;\{p_k\} ).\end{equation}
 \begin{Prop}\label{lemma2}
 	The total antisymmetrization $(\bar{\E}_N \bar{f})(x_1,...,x_N)$ can be described by the relation
 \begin{equation}\label{EI}
 	 (\bar{\E}_N \bar{f})(x_1,...,x_N)=\Delta(x_1,\dots ,x_N)\oint dz V'_-(z)V'_+(z)f(z;\{p_k\})
 \end{equation}
\end{Prop}
   Equivalently,
  \begin{equation*}\label{EI2}
 	{(\E_N f)(\{p_k\})=\oint dz V'_-(z)V'_+(z)f(z;\{p_k\})}.
 \end{equation*}
 
  Here on the RHS the function $f(z;\{p_k\})$ is a polynomial in $z$ and in $p_k$ depending on $(N-1)$ variables,
 while $V'_-(z)V'_+(z)f(z;\{p_k\})$ is a Laurent series in $z$ with coefficients being polynomials in $p_k$ depending on $N$ variables.  The integral on the right hand side counts the residue at infinity:
$$
\oint f(z)dz=f_{-1}\ \text{for} \ f(z)= \sum_i f_i z^i.
$$
The proof of Proposition \ref{lemma2} is based on the following statement.
 \begin{Lemma}\label{pr2} The following relation is valid
\begin{equation*} 
          x_1^k\Delta(x_2,\dots,x_N)-  x_2^k\Delta(x_1,x_3,\dots,x_N)+\dots+
           (-1)^{N+1}x_N^k\Delta(x_1,\dots,x_{N-1})=
           \end{equation*}
    \[
    =\begin{cases}
    \Delta(x_1,x_2,\dots,x_N)h_{k+1-N}(x_1,\dots,x_N) &  \text{  for  } k\ge N-1 \\
   0 & \text{  for  } 0\le k< N-1
    \end{cases},
  \]
  \end{Lemma}
Here $\displaystyle h_k(x_1,\dots,x_N)=\!\!\!\!\sum_{1\le i_1\le\cdots\le i_k\le N}\!\!\!x_{i_1}x_{i_2}\dots x_{i_N}$ are complete homogeneous symmetric polynomials.\\
\textbf{Proof of Lemma \ref{pr2}.}
Weyl formula for  Schur polynomials says
$$\displaystyle s_{(\lambda_1,\lambda_2,\dots,\lambda_N)}(x_1,x_2,\dots,x_N)=\det_{i,j=1\dots N}(x_i^{\lambda_j+N-j})/\Delta(x_1,\dots,x_N).$$ 
In particular, for $\displaystyle h_k(x_1,\dots,x_N)=s_{(k,0,0,\dots)}(x_1,\dots,x_N)$ we have 
\begin{equation}\label{hdet}
h_{k+1-N}(x_1,...,x_N)\Delta(x_1,x_2,\dots,x_N)=\det  \begin{pmatrix}
x_{1}^k & x_2^k & \dots  & x_{N}^k \\
x_{1}^{N-2} & x_2^{N-2} &\dots  & x_{N}^{N-2} \\
\vdots& \vdots & \ddots & \vdots \\
x_{1} & x_2  &\dots  & x_{N} \\
1 & 1& \dots  & 1 
\end{pmatrix}.
\end{equation}
For $0\le k< N-1$ the determinant in RHS of (\ref{hdet}) equals zero. The statement of lemma now follows from (\ref{hdet}) by the determinant expansion along the first row. 
See \cite[\$ 7]{Stanley}.  \hfill$\scriptstyle\blacksquare$
\smallskip

\noindent
\textbf{Proof of Proposition \ref{lemma2}}. Rewrite the relation (\ref{SK1}) in the form
  $$
\bar{f}(x_i;x_1,\dots, x_N)=(-1)^{i+1}\Delta(x_1,\dots x_{i-1},x_{i+1},\dots ,x_N) f^'(x_i;\{p_k\} ),
  $$
  where $f^'(x_i;\{p_k\}= V'_+(x_i) f(x_i;\{p_k\} )$ and $p_k$ depends on $N$ variables. The function $f^'(x_i;\{p_k\}) )$ is a polynomial in $x_i$ and $p_k$:
  $$
  f^'(x_i;\{p_k\} )=\sum_{l=0}^M x_i^l  f_l^'(\{p_k\})
  $$
  therefore we can realize antisymmetrization by each power of $x_i$ independently:
  \begin{equation*}
  \begin{split}
  (\bar{\E}_N \bar{f})(x_1,...,x_N)=& \sum_{l=0}^M f_l^'(\{p_k\})\left( x_1^l\Delta(x_2,x_3,\dots,x_N) \right. -x_2^l
  \Delta(x_1,x_3,\dots,x_N)+\dots \\ & \left. +(-1)^{N+1}x_N^l\Delta(x_1,x_2,\dots,x_{N-1})\right).
  \end{split}
  \end{equation*}
   Due to Lemma \ref{pr2}
    \begin{equation}\label{anti}
  (\bar{\E}_N \bar{f})(x_1,...,x_N)=\Delta(x_1,x_2,...,x_N)\sum_{l=N-1}^M f_l^'(\{p_k\})h_{l+1-N}(x_1,...,x_N)   .
  \end{equation}
Due to (\ref{V-f}) the formal integral 
\[
\oint dz V'_-(z)z^m= \begin{cases}h_{m+1-N}(\{p_n\}) &\text{  for  } m\ge N-1\\
0 & \text{  for  } 0\le m< N-1\end{cases},
\]
thus the integral $
\oint dz V'_-(z)f^'(z;\{p_k\})
$
gives the RHS of (\ref{anti}) divided by $\Delta(x_1,x_2,...,x_N)$. We then get  (\ref{EI}).\hfill$\scriptstyle\blacksquare$
\medskip

\textbf{6. } Let $f(x_i;\{p_k\})\in \C[x_i]\otimes \Sym[x_1,\dots x_{i-1},x_{i+1},\dots x_N$] and $\bar{f}(x_i;x_1,\dots, x_N)$ 
 be the corresponding element of the space $\C[x_i]\otimes \As[x_1,\dots x_{i-1},x_{i+1},\dots x_N]$:
   $$\bar{f}(x_i;x_1,\dots, x_N)=(-1)^{i+1}\Delta(x_1,\dots x_{i-1},x_{i+1},\dots ,x_N) f(x_i;\{p_k\} ).$$
 Define the operator
$$D_{i}^{(N)}: \C[x_i]\otimes \Sym[x_1,\dots x_{i-1},x_{i+1},\dots x_N]\to \C[x_i]\otimes \Sym[x_1,\dots x_{i-1},x_{i+1},\dots x_N]$$ 
by the relation
   \begin{equation}\label{De}
\begin{split}
D_{i}^{(N)}&f(x_i,\{p_k\})= x_i\frac{\partial}{\partial x_i}f(x_i,\{p_k\})+\\&
\beta x_i\oint dz\frac{ V'_-(z)V'_+(z)}{x_i-z}\left(V_-(z)V_+(z)f(x_i,\{p_k\})-V_-(x_i)V_+(x_i)f\left(z,\{p_k\}\right)\right).
\end{split}
\end{equation}  
  Then we formulate the following:
 \begin{Prop}\label{lemma3}
The action of the Dunkl operator $\bar{D}_i^{(N)}$ in the space of antisymmetric functions   $\C[x_i]$  $\otimes$     $\As[x_1,\dots,   x_{i-1},  x_{i+1},  \dots,x_N]$ can be expressed by the relation: 
\begin{equation}
\begin{split}
 \bar{D}_i^{(N)}\ \bar{f}(x_i;x_1,\dots, x_N)&=(-1)^{i+1}\overline{ D_{i}^{(N)}  f(x_i;\{p_k\} )}=\\&=(-1)^{i+1}\Delta(x_1,\dots x_{i-1},x_{i+1},\dots ,x_N) D_{i}^{(N)}  f(x_i;\{p_k\} ).  
 \end{split}
 \end{equation}

\end{Prop}
\begin{Proof}
 Firstly, we use the embedding $1\otimes\iota_{N,j}:\C[x_i]\otimes \Sym[x_1,\dots x_{i-1},x_{i+1},\dots x_N]\rightarrow \C[x_i]\otimes \C[x_j] \otimes \Sym[x_1,\dots x_{i-1},x_{i+1},\dots,x_{j-1},x_{j+1},\dots x_N] $ from  the proposition \ref{lemma1}:
 $$1\otimes\iota_{N,j}:f(x_i,\{p_n\})\rightarrow V_-(x_j)V_+(x_j)f(x_i,\{p_n\}) .$$ Then the operator $ \frac{x_i}{x_i-x_j}((1-K_{ij})$ can be written by the following formula 
 \begin{equation}
 \begin{split}
 &\frac{x_i}{x_i-x_j}(1-K_{ij})V_-(x_j)V_+(x_j)f(x_i,\{p_n\})=\\&= \frac{x_i}{x_i-x_j}(\left(V_-(x_j)V_+(x_j)f(x_i,\{p_n\})-V_-(x_i)V_+(x_i)f\left(x_j,\{p_n\}\right)\right).
 \end{split}
\end{equation}
Then we use the formula of total antisymmetrization from proposition \ref{lemma2}.
 \end{Proof}
 \medskip
 
 \textbf{7. } Here we present the formulas for antisymmetrization in a form which we will use in the Fock space limit.
 \begin{Remark}\label{r1}
 The formal integral $\oint dz V'_-(z)V'_+(z)f(z;\{p_k\})$ for the polynomial $f(z;\{p_k\})$ in $z$ can be rewritten as a complex integral
 \begin{equation}
  \frac{1}{(2\pi i)^2}\int_{z\around 0} dz\int_{u\around z}\!du  \frac{ V'_-(u)V'_+(u)f(z,\{p_k\})}{u-z}.
 \end{equation}

\end{Remark}
\begin{Remark}
 For $f(z;x_i;\{p_k\})$ with parameter $x_i$ the formal integral for antisymmetrization $\oint dz V'_-(z)V'_+(z)f(z; x_i; \{p_k\})$ can be rewritten as
  \begin{equation}
  \frac{1}{(2\pi i)^2}\int_{z\around 0, z\ll x_i} dz\int_{u\around z}\!du  \frac{ V'_-(u)V'_+(u)f(z;x_i;\{p_k\})}{u-z}.
 \end{equation}
\end{Remark}
 Here we choose the countour so as to avoid the singularity $z=x_i$. This is a rule for how to use the composition of Dunkl operators.
 
\textbf{8. }To obtain the Hamiltonians  
\begin{equation*}\notag
 \bar{H}^{(N)}_k=\sum_i ( \bar{D}^{(N)}_i)^k
\end{equation*}
we replace the outer sum by antisymmetrization operator $ \bar{\E}_N$ so that we get an expression which actually does not depend on $i$,
\begin{equation}\label{Hpol}
 \bar{H}^{(N)}_k=\bar{\E}_N(\bar{D}_i^{(N)})^k\bar{\iota}_{N,i}=\overline{\E_N({D}_i^{(N)})^k{\iota}_{N,i}}.
\end{equation}
The procedure is illustrated by the following diagram
\[  \begin{diagram}
\node{\As[x_1,\dots x_N]}
\arrow{e,t}{\bar{\iota}_{N,i}}
\node{\As[x_1,\dots x_{i-1},x_{i+1},\dots x_N]\otimes \C[x_i]}
\arrow{s,l}{(\bar{D}_i^{(N)})^k}
\\
\node[2]{\As[x_1,\dots x_{i-1},x_{i+1},\dots x_N]\otimes \C[x_i]}
\arrow{e,t}{ \bar{\E}_N}
\node{\As[x_1,\dots x_N]}
\end{diagram}.\]
The expressions for the first Hamiltonians $ {H}^{(N)}_k=\left(\E_N({D}_i^{(N)})^k{\iota}_{N,i}\right)$ are given below:
 $$
 H^{(N)}_0=\po,
 $$
 $$
 H^{(N)}_1=\sum_{n>0} n p_n\frac{\d}{\d p_n}+\left(1+2\beta\right)\frac{\po^2-\po}{2},
 $$
  \begin{equation*}
  \begin{gathered}
 H^{(N)}_2=\sum_{n,k>0} n k p_{n+k}\frac{\d}{\d p_n}\frac{\d}{\d p_k}+(1+\beta)\sum_{\substack{n,k> 0}} (n+k) p_n p_k\frac{\d}{\d p_{n+k}}-\beta\sum_{n>0} n^2 p_n\frac{\d}{\d p_n}\\
 -\left(1+2\beta\right)\sum_{n>0} n p_n\frac{\d}{\d p_n}+(3\beta+2) \po \sum_{n>0} n p_n\frac{\d}{\d p_n}
 \\+\frac{1}{6}(2\po^3-3\po^2+\po)+\frac{\beta}{6}(7\po^3-12\po^2+5\po)+\beta^2(\po^3-2\po^2+\po).
\end{gathered}
 \end{equation*}

\section{The limit}
\textbf{1.} Let  $\F=\Lambda[p_0]$ be the ring of symmetric functions \cite[II.2]{Mac} extended by the free variable $p_0$ , $\F=\C[p_0,p_1,\ldots]$.
The space $\F$ is an irreducible representation of the Heisenberg algebra, generated by the elements $p_n$ and $\dfrac{\partial}{\partial p_n}$ and can be regarded as a polynomial version of the Fock space. It contains the vacum vector
$|0\rangle$, such that $$\dfrac{\partial}{\partial p_n}|0\rangle=0,\qquad n=0,1, \ldots .$$
The dual vacuum vector $\langle 0|$ satisfies the condition
$$\langle 0| p_n=0, \qquad n=0,1,\ldots .$$

 Let $\Psi(z)$ and $\Psi^*(z)$ be vertex operators $\F\to\C[z,z^{-1}]]\otimes \F$,
 \begin{align}\label{psi}
 &\Psi(z)=z^{p_0}\exp\left(-\sum_{n>0}\frac{p_n}{n z^n}\right)\exp\left(\sum_{n\ge0}z^n\frac{\partial}{\partial p_n}\right),\\
 \label{psi*}
& \Psi^*(z)=z^{-p_0}\exp\left(\sum_{n>0}\frac{p_n}{n z^n}\right)\exp\left(-\sum_{n\ge0}z^n\frac{\partial}{\partial p_n}\right).
 \end{align}
 The following relations are valid:
 \begin{equation}\label{bu1}\begin{split}
 \Psi(z)\Psi(w)&=(w-z):\Psi(z)\Psi(w):
 \\
 \Psi(z)\Psi^*(w&)=\frac{1}{(w-z)}:\Psi(z)\Psi(w)^*:, \end{split}
 \end{equation}
 where $:\ :$ means bosonic normal ordering --- all operators $\frac{\d}{\d p_n} $ are moved to the right and operators $p_n$ are moved to the left.
 Operators (\ref{psi}) and (\ref{psi*}) satisfy the relations:
 	 \begin{equation*}
 	 	\frac{1}{2\pi i}\int_{z\around w} \Psi(w)\Psi^*(z)dz =\frac{1}{2\pi i}\int_{z\around w} \Psi^*(w)\Psi(z)dz=1.
 	 	\end{equation*}
 	 	\medskip

 \textbf{2.} Let $|v\rangle=f(p_0,p_1,...,p_k,...)|0\rangle\in \F$, where  $f(p_0,p_1,...,p_k,...)$ is a polynomial in $p_k$.
 Define the evaluation map $\pi_N:\F\rightarrow \As[x_1,\dots x_N]$ by the prescription
  \begin{equation}\label{pi}
 	 	\pi_N |v\rangle= \langle 0|\Psi(x_N)\cdots \Psi(x_1)|v\rangle.
 	 \end{equation}	
 	 The function $\pi_N |v\rangle$ is antisymmetric polynomial
 	 \begin{equation}\label{a1}
 	\pi_N |v\rangle= 	\prod_{i<j}(x_i-x_j)f(N,(x_1+\ldots+ x_N),...,(x_1^k+\cdots +x_N^k),...).
 	 	\end{equation}
 	 	Indeed, $\Psi(x_N)\cdots \Psi(x_1)=\prod_{i <j}(x_i-x_j):\Psi(x_N)\cdots \Psi(x_1):$ due to (\ref{bu1}).
 	 	 The operator $\prod_i\exp\left(\sum_{n\ge0}x_i^n\frac{\partial}{\partial p_n}\right)$ replaces every item $p_k$ in $f$  with $x_1^k+\cdots +x_n^k$, 
 	 	  $k=0,1,\ldots$, while 
 	 	$$\langle 0|\prod_i x_i^{p_0}\exp\left(-\sum_{n>0}\frac{p_n}{n x_i^n}\right)=\langle 0|.$$
 	 	\medskip

 	 	\textbf{3.} Similarly we define the map  $$\pi_{N-1,i}:z^{p_0}\C[z,z^{-1}]]\otimes\F\rightarrow \C[x_i, x_i^{-1}]]\otimes\As[x_1,\dots,x_{i-1},x_{i+1},\dots x_N]$$ as follows
 \begin{equation}\label{pii}
 		\pi_{N-1,i}:
 		 z^{p_0+k}\otimes |v\rangle \rightarrow 
 		(-1)^{i+1}\langle 0|\Psi(x_N)\cdots\Psi(x_{i+1})\Psi(x_{i-1})\cdots \Psi(x_1)x_i^{p_0+k}|v\rangle.
 	\end{equation}
Define the inclusion $\iota:\F\rightarrow z^{p_0}\C[z,z^{-1}]]\otimes\F$ by the relation $$\iota(|v\rangle)=\Psi(z)|v\rangle.$$
\begin{Lemma}\label{leminf1}
The following diagram is commutative:
\begin{equation}\label{diag1}
\begin{diagram}
\node{\F}
\arrow[2]{e,t}{\iota}
\arrow{s,l}{\pi_N}
\node[2]{z^{p_0}\C[z,z^{-1}]]\otimes\F} \arrow{s,r}{\pi_{N-1,i}} \\
\node{\As[x_1,\dots, x_N]}
\arrow[2]{e,b}{\bar{\iota}_{N,i}}
\node[2]{\C[x_i]\otimes\As[x_1,\dots,x_{i-1},x_{i+1},\dots x_N]} 
\end{diagram}
\end{equation}
\end{Lemma}
\begin{Proof}
Let us check the commutativity of the diagram (\ref{diag1}) for the element  $|v\rangle =f(p_0,p_1,...,p_k,...)|0\rangle\in\F$. The composition of $\pi_N$ and $\bar{\iota}_{N,i}$
defines the natural embedding of the antisymmetric polynomial
\begin{equation}\notag
\langle 0|\Psi(x_N)\cdots \Psi(x_1)|v\rangle
\end{equation} into the space  $\C[x_i]\otimes\As[x_1,\dots,x_{i-1},x_{i+1},\dots x_N]$, which is the expansion of the function in $x_1,\dots x_N$ over the variable $x_i$.
Applying the maps $\iota$ and $\pi_{N-1,i}$ we obtain the following relation
\begin{equation}\notag
\begin{split}
 \pi_{N-1,i}\iota |v\rangle&=\pi_{N-1,i}\Psi(z)|v\rangle=\\(-1)^{i+1}\langle 0|\Psi(x_N)\cdots\Psi(x_{i+1})\Psi(x_{i-1})\cdots \Psi(x_1)\Psi(x_i)|v\rangle&=\langle 0|\prod_{N\geq j\geq 1}\Psi(x_j)|v\rangle,
\end{split}
 \end{equation}
which coincides with natural embedding $\bar{\iota}_{N,i}$ of $\langle 0|\Psi(x_N)\cdots \Psi(x_1)|v\rangle$.
\end{Proof}
\medskip

\textbf{4. }Thus we have shown that for any $|v\rangle\in \F$ the element $\pi_{N-1,i}\iota(|v\rangle)\subset \C[x_i]\otimes\As[x_1,\dots,x_{i-1},x_{i+1},\dots x_N]$ is polynomial in $x_i$. Denote by $\WW$ the space
$$U=\cap_{{ N}}\pi_{N-1,i}^{-1}\left(\C[x_i]\otimes\As[x_1,\dots,x_{i-1},x_{i+1},\dots x_N]\right).$$ Due to Lemma \ref{leminf1} we have the inclusion $\iota(\F)\subset \WW$.

Define the map $\E:z^{p_0}\C[z,z^{-1}]]\otimes\F\rightarrow \F$ of antisymmetrization as follows
\begin{equation}\label{e}
  	\E F=\frac{1}{(2\pi i)^2}\int_{z\around 0} dz\int_{u\around z}\!du  \frac{\Psi^*(u)F(z)}{u-z}, 
  \end{equation}
  where $F(z)\in z^{p_0}\C[z,z^{-1}]]\otimes\F $. In other words 
  $$
  \E:z^{p_0+k}\otimes|v\rangle\rightarrow \frac{1}{(2\pi i)^2}\int_{z\around 0} dz\int_{u\around z}\!du  \frac{\Psi^*(u)z^{p_0+k}}{u-z}|v\rangle.
  $$
  \begin{Lemma}\label{leminf2}
The following diagram is commutative:
\begin{equation}\label{diag2}
\begin{diagram}
\node{z^{p_0}\C[z,z^{-1}]]\otimes\F\supset\WW}
\arrow[2]{e,t}{\E}
\arrow{s,l}{\pi_{N-1,i}}
\node[2]{\F} \arrow{s,r}{\pi_{N}} \\
\node{\C[x_i]\otimes\As[x_1,\dots,x_{i-1},x_{i+1},\dots x_N]}
\arrow[2]{e,b}{\bar{\E}_N}
\node[2]{\As[x_1,\dots, x_N]} 
\end{diagram}.
\end{equation}
\end{Lemma}
\begin{Proof} We can present any
 element in $\WW$  as a series $\sum_{k} z^{p_0+k}\otimes|v_{k}\rangle $.
We check the commutativity of the diagram (\ref{diag2}) for the element $z^{p_0+k}\otimes|v\rangle$, where $|v\rangle =f(p_0,p_1,...,p_k,...)|0\rangle$.
Following the definitions we obtain:
$$
\pi_{N-1,i}(z^{p_0+k}\otimes|v\rangle)=(-1)^{i+1}\langle 0|\Psi(x_N)\cdots\Psi(x_{i+1})\Psi(x_{i-1})\cdots \Psi(x_1)x_i^{p_0+k}f(p_0,p_1,\dots)|0\rangle.
$$
Thus $$
\pi_{N-1,i}(z^{p_0+k}\otimes|v\rangle)=\Delta(x_1,\dots x_{i-1},x_{i+1},\dots ,x_N)f(x_i;\{p_k\}),
$$
where
$
f(z;\{p_k\})=z^{k+N-1}f(N-1,p_1,p_2.\dots).
$
Using Proposition \ref{lemma2} we obtain
\begin{equation}\begin{split}\label{inf2}
\bar{\E}_N\pi_{N-1,i}& (z^{p_0+k}\otimes|v\rangle)=\\ & \frac{\Delta(x_1,\dots x_N)}{2 \pi i}\int_{z\around 0} dz V'_-(z)V'_+(z)z^{k+N-1}f(N-1,p_1,p_2,\dots)=
\\= &\langle 0|\Psi(x_N)\dots \Psi(x_1)\frac{1}{2\pi i}\int_{z\around 0} dz V'_-(z)V'_+(z)z^{k+N-1}f(N-1,p_1,p_2\dots)|0\rangle .
\end{split}
\end{equation}
Going by arrows  $\pi_{N} $ and $\E$ we get
\begin{equation*}\begin{split}
\pi_{N} \E( z^{p_0+k}& \otimes |v\rangle) =\\&\langle 0|\Psi(x_N)\cdots \Psi(x_1)\frac{1}{(2\pi i)^2}\int_{z\around 0} dz\int_{u\around z}\!du  \frac{\Psi^*(u)z^{p_0+k}}{u-z}f(p_0,p_1,\dots|0\rangle.
\end{split}\end{equation*}
To compare with the RHS of (\ref{inf2}) we use the following transformations:
\begin{align*}
\pi_{N} &\E(z^{p_0+k}\otimes|v\rangle)=\\ &\langle 0|\Psi(x_N)\cdots \Psi(x_1)\frac{1}{(2\pi i)^2}\int_{z\around 0} dz\int_{u\around z}\!du  \frac{V'_-(u)V'_+(u)e^{-\frac{\partial}{\partial p_0}}z^{p_0+k}}{u-z}f(p_0,p_1,\dots)|0\rangle=
\\&\langle 0|\Psi(x_N)\cdots \Psi(x_1)\frac{1}{(2\pi i)^2}\int_{z\around 0} dz\int_{u\around z}\!du  \frac{V'_-(u)V'_+(u)z^{k+N-1}}{u-z}f(p_0-1,p_1,p_2,\dots)|0\rangle=
\\
&\langle 0|\Psi(x_N)\dots \Psi(x_1)\frac{1}{2\pi i}\int_{z\around 0} dz V'_-(z)V'_+(z)z^{k+N-1}f(N-1,p_1,p_2,\dots)|0\rangle.
\end{align*}
Thus we prove the commutativity of the diagram (\ref{diag2}) for the element  $z^{p_0+k}\otimes|v\rangle$. For the sum  $\sum_{k} z^{p_0+k}\otimes|v_{k}\rangle $ we use the property of the space $\WW$, that its image by the projection 
$\pi_{N-1,1}$ is a finite sum.
\end{Proof}

\textbf{5. } Define the operator $D:\F\otimes\C[z,z^{-1}]]\rightarrow\F\otimes\C[z,z^{-1}]]$
\begin{equation}\label{Di}
{D} F(z)=z\frac{\partial}{\partial z}F(z)+
\beta \frac{1}{(2\pi i)^2}\int_{w\around 0} dw \int_{u\around w}\!\frac{du}{(u-w)}\frac{\Psi^*(u)}{\left(1-\frac{w}{z}\right)}\left(\Psi(w)F(z)-\Psi(z)F(w)\right).
\end{equation}
Due to Lemmas \ref{leminf1},\ref{leminf2} we get the following commutative diagram:
\begin{equation}\label{diagD}
\begin{diagram}
\node{\F\otimes\C[z,z^{-1}]]\supset \WW}
\arrow[2]{e,t}{\pi_{N-1,i}}
\arrow{s,l}{{D}}
\node[2]{\C[x_i]\otimes\As[x_1,\dots,x_{i-1},x_{i+1},\dots x_N]} \arrow{s,r}{\bar{D}_i^{(N)}} \\
\node{\F\otimes\C[z,z^{-1}]]\supset \WW}
\arrow[2]{e,b}{\pi_{N-1,i}}
\node[2]{\C[x_i]\otimes\As[x_1,\dots,x_{i-1},x_{i+1},\dots x_N]}
\end{diagram}.
\end{equation}
\textbf{6. }Define operators  $\H_k=\E {D}^k\iota:\F\to\F$ by the formula
\begin{equation} \label{ide}
\H_k:\F\xrightarrow{\iota}\WW\xrightarrow{{D}^k}\WW\xrightarrow{\E}\F.
\end{equation}
Due to (\ref{diagD}) we get the commutative diagram
\begin{equation}\label{diagH}
\begin{diagram}
\node{\F}
\arrow[2]{e,t}{\H_k}
\arrow{s,l}{\pi_{N}}
\node[2]{\F} \arrow{s,r}{\pi_{N}} \\
\node{\As[x_1,\dots ,x_N]}
\arrow[2]{e,b}{\bar{H}_k^{(N)}}
\node[2]{\As[x_1,\dots ,x_N]}
\end{diagram}.
\end{equation}

\begin{Prop}
 The operators $\H_k$ generate a commutative family of Hamiltonians of the limiting system.
\end{Prop}
\begin{Proof}
 For any $N$ operators $\bar{H}_k^{(N)}$ commute. Due to commutativity of (\ref{diagH}) and the fact that $
 \cap\, Ker(\pi_{N})= \varnothing
 $
 operators $\H_k$ commute as well.
\end{Proof}

We present the expression for the first Hamiltonians:
$$
 \H_0=p_0,
 $$
 $$
 \H_1=\sum_{n>0} n p_n\frac{\d}{\d p_n}+\left(1+2\beta\right)\frac{p_0^2-p_0}{2},
 $$
  \begin{equation*}
  \begin{gathered}
 \H_2=\sum_{n,k>0} n k p_{n+k}\frac{\d}{\d p_n}\frac{\d}{\d p_k}+(1+\beta)\sum_{\substack{n,k\ge 0 \\ n+k>0}} (n+k) p_n p_k\frac{\d}{\d p_{n+k}}\\-\beta\sum_{n>0} n^2 p_n\frac{\d}{\d p_n}
 -\left(1+2\beta\right)\sum_{n>0} n p_n\frac{\d}{\d p_n}
 +\beta p_0 \sum_{n>0} n p_n\frac{\d}{\d p_n}+\\ +\frac{1}{6}(2p_0^3-3p_0^2+p_0)+\frac{\beta}{6}(7p_0^3-12p_0^2+5p_0)+\beta^2(p_0^3-2p_0^2+p_0).
\end{gathered}
 \end{equation*}
 The limiting expression $\H$ corresponding to (\ref{H})  can be expressed by the formula similar to (\ref{prH}):
 \begin{equation*}
  \begin{gathered}
\H=\H_2-2\beta (p_0-1)\H_1 +\beta^2p_0(p_0-1)^2=\\
=\sum_{n,k>0} n k p_{n+k}\frac{\d}{\d p_n}\frac{\d}{\d p_k}+(1+\beta)\sum_{\substack{n>0,k\ge 0}} (n+k) p_n p_k\frac{\d}{\d p_{n+k}}-\beta\sum_{n>0} n^2 p_n\frac{\d}{\d p_n}\\
 +(p_0-1) \sum_{n>0} n p_n\frac{\d}{\d p_n} +\frac{1}{6}(2p_0^3-3p_0^2+p_0)+\frac{\beta}{6}p_0(p_0^2-1).
\end{gathered}
 \end{equation*}
The Hamiltonian $\H+\H_1$ with shift $\beta\rightarrow (\beta-1)$ coincides with the bosonic limiting expression \cite{NazSk, SerVes2} by putting $p_0=0$.

{\textbf{ 7 Comments.}} The space $\Lambda$ of symmetric functions can be realized either as the projective limit of the 
rings of symmetric polynomials in $N$ variables, or the projective limit of the spaces of antisymmetric polynomials in $N$ variables. The latter means the commutativity of the diagrams 
\begin{equation*}
\begin{diagram}
\node[2]{\Lambda}
\arrow[1]{se,t}{\alpha_{N+1}}
\arrow{sw,l}{\alpha_{N}}
\\
\node{\As_N}
\node[2]{\As_{N+1}}
\arrow[2]{w,b}{\lambda_N}
\end{diagram},
\end{equation*}
where $\As_N=\As[x_1,x_2, \dots , x_{N}] $, 
\begin{align*}
&\lambda_N: \bar{f}(x_1,\dots x_N,x_{N+1})\mapsto \bar{f}(x_1,\dots x_N,0)\prod_{i=1}^N x_i^{-1} ,\qquad 
\text{and}\\ & \alpha_N: f(p_1,\cdots p_N) \mapsto
	\prod_{i<j}(x_i-x_j)f((x_1+\ldots+ x_N),...,(x_1^k+\cdots +x_N^k),...).
\end{align*}
 The space $\F$ is not a projective limit of the spaces $\As_N$ due to the presence of $p_0$ which breaks
 the commutativity of analogous diagram for $\F$ with $\alpha_N$ replaced by maps
  $\pi_{N}$.  
On the other hand, CS Hamiltonians $\bar{H}_k$ theirselves do not compose the projective system since
$\lambda_N\bar{H}_k^{(N+1)}\neq\bar{H}_k^{(N)}\lambda_N
$. 
However, the Hamiltonians $\bar{H}_k^{(N)}$ written in form (\ref{Hpol}) are compatible with  maps $\lambda_N$, if we replace each occurrence of $N$ in $\bar{H}_k^{(N)}$
to $N+1$ in $\bar{H}_k^{(N+1)}$. Moreover, each finite Hamiltonian can be  restored from its limit by formal replacement of each occurrence of $p_0$ by operator of multiplication on the number $N$ of particles.

This correspondence hints the form of corrections in Hamiltonians to form a projective system: substract terms containing $p_0$ in the limit expression. Here are examples of corrections for the first Hamiltonians:
\begin{equation*}
  \begin{split}
&\bar{H}_{pr,1}^{(N)}=\bar{H}_1^{(N)}-(1+2\beta)\frac{N^2-N}{2},
\\&
\bar{H}_{pr,2}^{(N)}=\bar{H}_2^{(N)}-3\beta N \bar{H}_{pr,1}^{(N)}
\\&
-\frac{1}{6}(2\po^3-3\po^2+\po)-\frac{\beta}{6}(7\po^3-12\po^2+5\po)-\beta^2(\po^3-2\po^2+\po).
\end{split}
\end{equation*}
\section{Realization in the Fock space}
{\bf 1}. The constructed above Hamiltonians  form a commutative family of operators in the space $\F$. Moreover, they commute inside the Heisenberg algebra and thus can be used as well in its other representations, for instance, 
in the bosonic Fock space $\Fo$. In this section we show how to realize the limit in the bosonic Fock space, the key point is to define the analog of projection $\pi_N$. The formulas for the Hamiltonians remains the same.

The bosonic Fock space is usually defined as a free commutative algebra $\C[q,p_1,p_2,\dots]$  on varibles $p_k$ and $q$. Define the vacuum vector  $|0\rangle$  and a dual vacuum $\langle0|$ of the bosonic Fock space $\Fo$:
 \begin{equation*} \frac{\partial}{\partial p_n}|0\rangle=0,\quad n\geq 1\qquad \langle0|p_n=0,\qquad n\geq 0,\quad \langle0|p_0=p_0|0\rangle=0 . \end{equation*}
Denote by $\langle n|$  and $|n\rangle$ the following vectors:
 \begin{equation*}  |n\rangle=e^{-n\frac{\partial}{\partial p_0}}|0\rangle=q^{-\ n}|0\rangle,\qquad \langle n|=\langle0|q^{n}.
 	\end{equation*}
 	These vectors are biorthogonal $\langle n|m\rangle= \delta_{n,m}$ and have the following properties
 	\begin{equation*} \langle n|p_0=n\langle n|,\quad p_0|n\rangle=n|n\rangle.
 \end{equation*}
  Any vector in space $\Fo$ can be presented as a linear combination of such vectors $|v\rangle=f(p_1,...,p_k,...)|c\rangle$, where $f(p_1,...,p_k,...)$ is a polynomial in $p_k$ and $c$ is so called charge of $|v\rangle$ and we denote it by $p_0(v)$. Denote by $\Fo_c$ 
  the linear span of vectors with charge $c$, then $\Fo$ is graded according to the charge $\Fo=\oplus_{c\in \Z}\Fo_c$.

 Define the projection $\tilde{\pi}_N:\Fo\rightarrow \As[x_1,\dots x_N]$ by the prescription
 \begin{equation}\label{tildepi}
 \tilde{\pi}_N |v\rangle= \langle 0|\Psi(x_N)\cdots \Psi(x_1)|v\rangle.
 \end{equation}
 Due to biorthogonality $\langle n|m\rangle= \delta_{n,m}$ and fact that product $\Psi(x_N)\cdots \Psi(x_1)$ contains $q^N$ we have $\tilde{\pi}_N(\Fo_c)=0$  for $c\neq N$. Thus for $|v\rangle=f(p_1,...,p_k,...)|c\rangle$ we have
 \[
 \tilde{\pi}_N |v\rangle=\begin{cases}
                        \prod_{i<j}(x_i-x_j)f((x_1+\ldots+ x_N),...,(x_1^k+\cdots +x_N^k),...)\ \ \text{for } p_0(v)=N \\
                          0\ \ \ \text{for }  p_0(v)\neq N
                         \end{cases}.
 \]
 Similarly we define the map  $$\tilde{\pi}_{N-1,i}:z^{p_0}\C[z,z^{-1}]]\otimes\Fo\rightarrow \C[x_i, x_i^{-1}]]\otimes\As[x_1,\dots,x_{i-1},x_{i+1},\dots x_N]$$ as follows
 \begin{equation}\notag
 \begin{split}
 		\tilde{\pi}_{N-1,i}:& z^{p_0+k}\otimes |v\rangle \rightarrow \\ &(-1)^{i+1}\langle 0|\Psi(x_N)\cdots\Psi(x_{i+1})\Psi(x_{i-1})\cdots \Psi(x_1)x_i^{p_0+k}|v\rangle.
 	\end{split}\end{equation}
 	Due to the same arguments $\tilde{\pi}_{N-1,i}(z^{p_0}\C[z,z^{-1}]]\otimes\Fo_c)=0$ if $c\neq N$. Then we have the analogous commutativity as in Lemma \ref{leminf1} for $\Fo$ and $\tilde{\pi}_N$ instead of $\F$ and $\pi_N$, 
 	which is nontrivial only for the sector $\Fo_N$, the proof remains the same.
 	 Denote by $\tilde{\WW}_N\subset z^{p_0}\C[z,z^{-1}]]\otimes\Fo$ the space $$\tilde{\WW}_N=\tilde{\pi}_{N-1,i}^{-1}\left(\C[x_i]\otimes\As[x_1,\dots,x_{i-1},x_{i+1},\dots x_N]\right).$$ 
 	 We have the inclusion $\iota(\Fo_N)\subset \tilde{\WW}_N$. The analogous commutativity as in Lemma \ref{leminf2} holds:
 	 \begin{equation}\label{diag222}
\begin{diagram}
\node{z^{p_0}\C[z,z^{-1}]]\otimes\Fo_{N-1}\supset\tilde{\WW}_N}
\arrow[2]{e,t}{\E}
\arrow{s,l}{\tilde{\pi}_{N-1,i}}
\node[2]{\Fo_{N}} \arrow{s,r}{\tilde{\pi}_{N}} \\
\node{\C[x_i]\otimes\As[x_1,\dots,x_{i-1},x_{i+1},\dots x_N]}
\arrow[2]{e,b}{\bar{\E}_N}
\node[2]{\As[x_1,\dots, x_N]} 
\end{diagram}.
\end{equation}
The proof may be reproduced as in Lemma \ref{leminf2} changing each occurrence of $p_0$ by $N-1$ due to $\tilde{\WW}_N\in z^{p_0}\C[z,z^{-1}]]\otimes\Fo_{N-1}$. Thus we have the commutative diagram for the Dunkl operators 
which is nontrivial for the $N$-th sector of the Fock space $\Fo_N$:
\begin{equation}\label{diagD2}
\begin{diagram}
\node{\Fo_{N-1}\otimes\C[z,z^{-1}]]\supset \tilde{\WW}_N}
\arrow[2]{e,t}{\tilde{\pi}_{N-1,i}}
\arrow{s,l}{{D}}
\node[2]{\C[x_i]\otimes\As[x_1,\dots,x_{i-1},x_{i+1},\dots x_N]} \arrow{s,r}{\bar{D}_i^{(N)}} \\
\node{\Fo_{N-1}\otimes\C[z,z^{-1}]]\supset \tilde{\WW}_N}
\arrow[2]{e,b}{\tilde{\pi}_{N-1,i}}
\node[2]{\C[x_i]\otimes\As[x_1,\dots,x_{i-1},x_{i+1},\dots x_N]}
\end{diagram}.
\end{equation}
On the other sectors of the Fock space (\ref{diagD2}) holds due to $\tilde{\pi}_{N-1,i}$ projects all to zero.
We arrive to the following 
\begin{Prop} The Hamiltonians $\H_k:\Fo\to\Fo$ are the pullback of Hamiltonians $\bar{H}_k^{(N)}$ with respect to
		the maps $\tilde{\pi}_{N}$.	
	\end{Prop}
In other words, the Hamiltonians (\ref{ide}) obey the commutative diagram
\begin{equation}\label{diagH2}
\begin{diagram}
\node{\Fo}
\arrow[2]{e,t}{\H_k}
\arrow{s,l}{\tilde{\pi}_{N}}
\node[2]{\Fo} \arrow{s,r}{\tilde{\pi}_{N}} \\
\node{\As[x_1,\dots ,x_N]}
\arrow[2]{e,b}{\bar{H}_k^{(N)}}
\node[2]{\As[x_1,\dots ,x_N]}
\end{diagram}.
\end{equation}

{\bf 2}. Now we want to describe the construction in the fermionic Fock space realized as semi-infinite wedges and present the projection analogous to $\tilde{\pi}_N$. We introduce the Clifford algebra generated by fermions $\psi_k,\psi_k^*$ for $k\in \Z$ with  anti-commutation relations
 $$
 \psi_i\psi_j+\psi_j\psi_i=\psi_i^*\psi_j^*+\psi_j^*\psi_i^*=0,
$$
$$
 \psi_i\psi_j^*+\psi_j^*\psi_i=\delta_{ij}.
 $$
 The fermionic Fock space $\Fo$ can be defined as a representation of the Clifford algebra, where the vacuum vetor $|0\rangle$ is defined as follows:
 \begin{equation}\label{vacuum}
\psi_n|0\rangle=0 \ \ n\ge 0,\ \
\psi_n^*|0\rangle=0 \ \ n< 0.
\end{equation}
 According to (\ref{vacuum}) the fermionic normal ordering ${\vdots \ \vdots}$ is defined as follows:
 \[
 {\vdots\psi_i^*\psi_j\vdots}=\begin{cases}
\psi_i^*\psi_j, \ \ j\ge 0\\
-\psi_j\psi_i^*,\ \ j<0
\end{cases}.
 \]
 In other words all annihilation operators are moved to the right and all creation operators are moved to the left taking into account that the factor $(-1)$ appears after exchanging neighboring fermionic operators.
Any wedge in the space $\Lambda^{\frac{\infty}{2}}(\mathbb{C}[z,z^{-1}])$ can be obtained by acting of fermionic operators on the vacuum state 
\begin{equation}\label{f}
{\vdots\psi_{k_1}\psi_{k_2}\dots\psi_{k_n}\psi_{l_1}^*\psi_{l_2}^*\dots\psi_{l_m}^*\vdots}|0\rangle.
\end{equation}
A charge of element (\ref{f}) can be defined as $m-n$. We introduce the shifted vacuum $|c\rangle$ 
\begin{equation}\notag
|c\rangle= 
    \begin{cases}
\psi^*_{c-1}\dots\psi^*_1\psi^*_0|0\rangle &\ \  c>0\\
\psi_{c}\dots\psi_{-2}\psi_{-1}|0\rangle    &\ \  c<0
    \end{cases}.
\end{equation}
In $\Fo$ we can choose a basis $|\lambda,c\rangle$ parameterized by partition $\lambda=(\lambda_1,\lambda_2,\dots,\lambda_{n})$: 
\begin{equation}\label{lamc}
 |\lambda,c\rangle=\psi^*_{\lambda_1-1}\psi^*_{\lambda_2-2}\dots\psi^*_{\lambda_n-n}|c-n\rangle.
\end{equation} 
For fixed $c$ vectors $|\lambda,c\rangle$ generate the $c$-th sector $\Fo_c$ of the fermionic Fock space as a vector space.

The fermionic Fock space  admits a presentation $\Fo\cong \Lambda^{\frac{\infty}{2}}(\mathbb{C}[z,z^{-1}])$ in ``semi-infinite wedges'':
$$
z^{k_1}\wedge z^{k_2}\wedge \dots \wedge z^{k_{m}}\wedge \dots, \ \ k_1>k_2>\dots>k_{m}>\dots, \ \ k_{n+1}=k_n-1 \ \text{all}\ n>N,
$$
which form a basis of $\Fo$. 
The vacuum state $|0\rangle$ corresponds to 
$$
|0\rangle=z^{-1}\wedge z^{-2}\wedge z^{-3}\wedge z^{-4}\wedge \dots
$$ An action of fermionic operators on the wedge $v$ is presented by formulas:
$$
\psi_n(v)=\frac{\partial}{\partial z^n} v,\ \ 
\psi_n^*(v)=z^n\wedge v.
$$
Note that the element $z^n$ is added by $\psi^*_n$ at the very beginning of the sequence, so the permutiaion with other elements may produce a sign.  The symbol $\frac{\partial}{\partial z^n}$ means 
that if the wedge $v=z^n\wedge w$ then 
$$\frac{\partial}{\partial z^n}( z^n\wedge  w)=w.$$
The shifted vacuum is given by 
$$
|c\rangle=z^{c-1}\wedge z^{c-2}\wedge z^{c-3}\wedge z^{c-4}\dots.
$$
and  $|\lambda,c\rangle$ from (\ref{lamc})
\begin{equation}\notag
 |\lambda,c\rangle=z^{\lambda_1+c-1}\wedge z^{\lambda_2+c-2}\wedge\dots\wedge z^{\lambda_k+c-k}\wedge\dots\wedge z^{\lambda_n+c-n}\wedge z^{-n-1+c}\wedge z^{-n-2+c}\dots.
\end{equation}

Define the space  $\Lambda^{N}(\mathbb{C}[z,z^{-1}])$ of finite wedge $z_{1}^{k_1}\wedge z_2^{k_2}\wedge \dots \wedge z_N^{k_N}$  with $N$ elements. It can be identified with the antisymmetric function  
$\Lambda^{N}(\mathbb{C}[z,z^{-1}])\simeq \As[z_1^{\pm 1},\dots,z_N^{\pm 1}]$:
\begin{equation}\label{ism}
z_{1}^{k_1}\wedge z_2^{k_2}\wedge \dots \wedge z_N^{k_N}\Longleftrightarrow \displaystyle Alt(z_{1}^{k_1},\dots,z_N^{k_N})=\det_{i,j=1\dots N}{z_i^{k_j}}.
\end{equation}
For wedge $
v=z^{k_1}\wedge z^{k_2}\wedge \dots \wedge z^{k_i}\wedge \dots \in \Lambda^{\frac{\infty}{2}}(\mathbb{C}[z,z^{-1}])
$ denote by $p_0(v)$ the charge of $v$. We can define the embedding 
$\omega_N:\Lambda^{\frac{\infty}{2}}(\mathbb{C}[z,z^{-1}])\rightarrow \Lambda^{N}(\mathbb{C}[z,z^{-1}])$ :
\begin{equation}\label{omega}
\omega_N(v)= 
    \begin{cases}
 z_{1}^{k_1}\wedge z_2^{k_2}\wedge \dots \wedge z_N^{k_N} &  \text{  if  } p_0(v)=N\\
   0 & \text{  if  } p_0(v)\neq N
    \end{cases}
\end{equation}
that simply keep only the first $N$ elements in wedge $v$ if its charge equals $N$.
For a partition $\lambda=(\lambda_1,\lambda_2,\dots,\lambda_{n})$ we have
\begin{equation}\notag
\omega_N(|\lambda,c\rangle)= 
    \begin{cases}
 z_{1}^{\lambda_1+N-1}\wedge z_2^{\lambda_2+N-2}\wedge \dots \wedge z_N^{\lambda_N} &  \text{  if  } c=N\\
   0 & \text{  if  } c\neq N
    \end{cases},
\end{equation}
where we put $\lambda_i=0$ for $i > n$. Due to the isomorphism (\ref{ism}) we obtain
$$
\omega_N(|\lambda,N\rangle)\simeq \det_{i,j=1\dots N}{z_i^{\lambda_j+N-j}}=\prod_{i<j}\left({z_i}-{z_j}\right)s_\lambda(z_1,z_2,\dots,z_N),
$$
where $s_\lambda(z_1,z_2,\dots,z_N)$ is a Schur polynomial.
Define operators
 \begin{equation}\label{bos}
a_n=\sum_j\vdots\psi_j^*\psi_{j+n}\vdots.
\end{equation}
It can be checked that they commute as bosonic operators
$$
[a_k,a_l]=k\delta_{k+l,0}.
$$
Define the operator $Q$ with the following commutation relations 
$$
[a_n,Q]=\delta_{0,n}.
$$
The operator $e^Q$ is an operator which shifts the charge of the fermionic vector :
$$
e^Q \psi_n e^{-Q}=\psi_{n+1}, \ \ e^Q \psi_n^* e^{-Q}=\psi^*_{n+1}.
$$
Define the fermion field $\psi(x)=\sum_k \psi_k x^k$ and $\psi^*(x)=\sum_k\psi_k^* x^{-k-1}$ with
$$
\psi^*(x)\psi(x')=\frac{1}{x-x'}+\text{reg}.
$$
The boson fermion correspondence is given by the formula (\ref{bos}) and the following relations:
\begin{equation}\label{fer}
 \psi(x)= :x^{a_0}e^{-Q}\exp\left(-\sum_{n>0}\frac{a_{-n}}{n x^n}\right)\exp\left(\sum_{n>0} \frac{a_{n}}{n}x^n\right):
\end{equation}
$$
\psi^*(x)=:x^{-a_0}e^{Q}\exp\left(\sum_{n>0}\frac{a_{-n}}{n x^n}\right)\exp\left(-\sum_{n>0} \frac{a_{n}}{n}x^n\right):
$$
This corresponds with the notations given at the beginning of this paragraph where we put:
$$
a_{-n} = p_n,\ \ a_n=n\frac{\d}{\d p_n}\ \text{for } n>0 ,
$$
$$
a_0=p_0, \ Q=-\frac{\d}{\d p_0}.
$$
and with notations of vertex operators (\ref{psi}) which are representation of $\psi(z)$ and $\psi^*(z)$.
Due to the boson-fermion correspondence we formulate the following
\begin{Prop} 
 The diagram (\ref{diagbf1}) is commutative for $N>0$. 
\begin{equation}\label{diagbf1}
\begin{diagram}
\node{\Fo^{bos}}
\arrow[2]{e,t}{}
\arrow{s,l}{\tilde{\pi}_{N}}
\node[2]{\Fo^{fer}} \arrow{s,r}{\omega_{N}}
\arrow[2]{w,t}{}\\
\node{\As[x_1,\dots ,x_N]}
\arrow[2]{e,b}{}
\node[2]{\Lambda^{N}(\mathbb{C}[z,z^{-1}])}
\arrow[2]{w,b}{}
\end{diagram}
\end{equation}
\end{Prop}
Here the upper isomorphism is the boson-fermion correspondence (\ref{bos}), (\ref{fer}). The lower isomorphism is given by (\ref{ism}).
\begin{Proof}
 Consider a vector $|\lambda,c\rangle\in\Fo_c^{fer}$ for a partition $\lambda=(\lambda_1,\lambda_2,\dots,\lambda_{n})$. We have shown that 
 \[
\omega_N(|\lambda,c\rangle)=\begin{cases} \prod_{i<j}\left({z_i}-{z_j}\right)s_\lambda(z_1,z_2,\dots,z_N)\ \ \text{if } c=N
\\0 \ \ \text{if } c\neq N                         
                            \end{cases}
,
\]
for $n\le N$. One can show \cite{AZ} that boson-fermion correspondence implies $|\lambda,c\rangle\simeq s_\lambda({\bf p})|c\rangle$, where $s_\lambda({\bf p})$ is a Schur polynomial in terms of $p_k$. 
Applying (\ref{tildepi}) to  $s_\lambda({\bf p})|c\rangle$ we obtain
 \[
\tilde{\pi}_N(s_\lambda({\bf p})|c\rangle)=\begin{cases} \prod_{i<j}\left({z_i}-{z_j}\right)s_\lambda(z_1,z_2,\dots,z_N)\ \ \text{if } c=N
\\0 \ \ \text{if } c\neq N                         
                            \end{cases}.
\]
\end{Proof}

\section*{Acknowledgements} 
The authors are grateful to M.L. Nazarov and E.K.Sklyanin for fruitful
discussions on the subject of the paper. The research by M.M. was carried out within the HSE University Basic Research Program
                                                     and funded jointly by the Russian Academic Excellence Project '5-100'. It was also supported in part by the Simons Foundation.
                                                     S.K appreciates the support of Russian Sience Foundation grant, project 16-11-10316 used for the proofs of propositions 2,3 section 3.


\end{document}